\documentclass[journal]{IEEEtran}
\usepackage[edges]{forest}
\usetikzlibrary{arrows.meta}
\usepackage{graphicx,times,amsmath,cite,enumerate}
\usepackage{kantlipsum}
\usepackage{threeparttable}
\usepackage{epsfig}
\usepackage{amssymb}
\usepackage{latexsym}
\usepackage{psfrag}
\usepackage{setspace}
\usepackage{color}
\usepackage{verbatim}
\usepackage{amsthm}
\usepackage{footnote}
\usepackage{textcase}
\usepackage{array}
\newcolumntype{P}[1]{>{\centering\arraybackslash}p{#1}}
\usepackage{float} 
\usepackage{multirow}
\usepackage{textcomp}
\usepackage{forest}
\PassOptionsToPackage{hyphens}{url}
\usepackage{url}
\usepackage{physics}
\usepackage{hyperref}
\usepackage{etoolbox}
\usepackage[edges]{forest}
\usepackage[ruled,vlined]{algorithm2e}
\usepackage{siunitx}

\usepackage{subfigure}
\usepackage{physics}
\usepackage{tikz}
\usetikzlibrary{automata, positioning}
\usepackage{nomencl}
\usepackage{float}
\usepackage{tabularx,booktabs}
\usepackage{url}

\title{Impact of the Inflation Reduction Act and Carbon Capture on Transportation Electrification for a Net-Zero Western U.S. Grid}

\author{
\IEEEauthorblockN{Samrat Acharya\IEEEauthorrefmark{1}, Malini Ghosal\IEEEauthorrefmark{1}, Travis Thurber\IEEEauthorrefmark{1}, Ying Zhang\IEEEauthorrefmark{1},  Casey D. Burleyson\IEEEauthorrefmark{1}, and Nathalie Voisin\IEEEauthorrefmark{1}\IEEEauthorrefmark{2} }\\
\IEEEauthorblockA{\IEEEauthorrefmark{1} \textit{Pacific Northwest National Laboratory, Richland, WA 99354, USA}} \\
\IEEEauthorblockA{\IEEEauthorrefmark{2} \textit{University of Washington, Seattle, WA 98195, USA}} \\
\{samrat.acharya\}@pnnl.gov \vspace{-0pt}
}

\IEEEoverridecommandlockouts

\begin{document}
\bstctlcite{IEEEexample:BSTcontrol} 

\maketitle
\begin{abstract}
The electrification of transportation is critical to mitigate Greenhouse Gas (GHG) emissions. The United States (U.S.) government's Inflation Reduction Act (IRA) of 2022 introduces policies to promote the electrification of transportation. In addition to electrifying transportation, clean energy technologies such as Carbon Capture and Storage (CCS) may play a major role in achieving a net-zero energy system. Utilizing scenarios simulated by the U.S. version of the Global Change Analysis Model (GCAM-USA), we analyze the individual and compound contributions of the IRA and CCS to reach a clean U.S. grid by 2035 and net-zero GHG emissions by 2050. We analyze the contributions based on three metrics: i) transportation electrification rate, ii) transportation fuel mix, and iii) spatio-temporal charging loads. Our findings indicate that the IRA significantly accelerates transportation electrification in the near-term (until 2035). In contrast, CCS technologies, by enabling the continued use of internal combustion vehicles while still advancing torward net-zero, potentially suppresses the rate of transportation electrification in the long-term. This study underscores how policy and technology innovation can interact and sensitivity studies with different combination are essential to characterize the potential contributions of each to  the transportation electrification. 

\end{abstract}


\section{Introduction}
\label{sec:intro}
 \IEEEPARstart{T}he urgency to mitigate climate change has prompted governments and concerned authorities worldwide to propose ambitious decarbonization targets aimed at reducing Greenhouse Gas (GHG) emissions. A myriad of factors influence the potential pathways to meet those targets including projected advancements in technology, socio-economic considerations, and packages of incentives and policies that combine those considerations\cite{ou2021can}. The inherent uncertainty surrounding these factors and their combinations lead to the existence of multiple potential pathways to achieve a given decarbonization objective\cite{bistline2023emissions}. In this paper, we scrutinize specifically the pathways to decarbonize the transportation sector, a major GHG emitter in the United States (U.S.) \cite{us_ghg}. We quantify for the first time across the whole vehicle fleet the coupled influences of the U.S. Inflation Reduction Act (IRA) and Carbon Capture and Storage (CCS) technologies on transportation electrification rates. This combination is important because the IRA policies tend to incentivize the electrification of transportation sector while CCS technologies are expected to provide relief in that same electrification \cite{kalkuhl2015role}. We quantify a range of uncertainties in key transportation sector metrics to inform power system operators and utilities in their long-term planning for resource adequacy and Electric Vehicle (EV) charging infrastructure planners and investors for meeting the decarbonization objectives.

The U.S. government's IRA, enacted in 2022, stands as one of the most substantial investments aimed at combating climate change and steering the economy towards net-zero emissions by 2050 \cite{IRA}. This comprehensive initiative includes incentives targeting various aspects of clean energy production as well as the domestic manufacturing of technologies crucial for reducing carbon emissions such as batteries and EVs. Specifically, within the transportation sector, the IRA encompasses provisions for promoting clean fuel vehicles, incentivizing the establishment of charging infrastructure, and encouraging domestic production of clean vehicles and their components. Section~\ref{sec:IRA_transportation_gcam} will delve into the specifics of the clean transportation measures outlined within the IRA. Given the potential interactions of IRA policies with CCS technologies, a sensitivity around the IRA provides a benchmark for isolating the individual versus impacted contributions to the electrification of transportation sector. The IRA is set to expire in 2032 while the net zero economy goal is expected to be achieved by 2050. It is therefore strategic to isolate the impact of the policy through 2050 which can provide information to planners such as the value for early investors (by 2032) versus late investors (past 2032).

 \begin{table}[!b]
\centering
\begin{tabular}{ |p{0.46\textwidth}| }
 \hline
\rule{0pt}{2.5ex}  This is a preprint. It's complete copyright version will be available on the publisher's website after publication. \\
 \hline
\end{tabular}
\end{table}

CCS are necessary technologies to combat climate change \cite{ccs_importance}. For instance, various climate change assessments such as the Intergovernmental Panel on Climate Chang, the International Energy Agency, and the UK’s Committee on Climate Change indicate that the GHG emission reduction targets set in the 2015 Paris Agreement \cite{paris_aggrement} cannot be met without CCS \cite{ccs_importance}. CCS technologies primarily isolate and capture CO$_2$ from industrial emissions and from atmosphere (e.g., direct air capture), storing it underground or sequestering it. These technologies are only beginning to be deployed \cite{gcam_docuv7, gcam_docuv6}.
For example, only 15 CCS facilities were deployed in the U.S. as of 2023, accounting for $0.4\%$ of nation-wide CO$_2$ emissions \cite{ccs_us_congressional_budget_office}. Furthermore, the survey in \cite{pianta2021carbon} finds U.S. residents with low awareness and support on CCS, primarily because of their rising cost and deployments near residential areas. However, several studies envisage the significant deployment of CCS in the next decade \cite{ccs_us_congressional_budget_office, ausfelder2020special}. Thus, understanding the contribution of CCS technologies to various potential decarbonization scenarios might help develop incentives to its adoption in line with its contribution to the overarching decarbonization goals.

In this study, we use the U.S. version of the Global Change Analysis Model (GCAM-USA), an open source multi-sectoral economics model \cite{calvin2019gcam}, to develop scenarios that reflect the IRA policies and potential widespread adoption of CCS technologies. These scenarios incorporate three U.S. decarbonization goals; i) to reduce nationwide GHG emissions by 50\% in 2030 relative to 2005 levels, ii) to achieve a clean U.S. power grid by 2035, and iii) to achieve a net-zero nationwide economy by 2050. Details on GCAM-USA and the scenarios explored are provided in Section~\ref{sec:gcam}. To asses the impact of IRA policies on transportation, we study two scenarios that reflect combinations of scenarios with and without the IRA policies for achieving a clean U.S. grid by 2035. Similarly, to asses the impact of CCS technologies on transportation, we use two scenarios for achieving net-zero GHG emissions across the U.S. economy by 2050. Notably, we focus our analysis on IRA policies from 2025 to 2035 and CCS technologies during 2035 to 2050 because of the separate IRA functional period and CCS deployment time scale. Additionally, although IRA policies conclude in 2032, we extend our analysis through 2035 to align with GCAM-USA's five-year scenario modeling increments. However, we also study the complementary relationship of the IRA policies and CCS technologies to achieve the 2050 net-zero economy goal. We focus our analysis on the Western U.S. Interconnection. We use the spatio-temporal downscaling approach developed for transportation electric loads in \cite{acharya2023weather} to analyze the impact of IRA policies and CCS technologies on transportation electrification across the Balancing Authorities (BAs) in the Western U.S. Interconnection. Although our analysis is focused on the Western U.S. Interconnection, the study's methodology can be applied to other regions in the U.S., since GCAM-USA models all the states. To this end, we make the following unique contributions:
\begin{enumerate}
    \item We analyze the relative impacts of the IRA policies and CCS technologies on transportation electrification by transportation fuel types, vehicle types, electrification rates, states, BAs, and seasons.
    
    \item We provide the spatio-temporal transportation electrification dataset by BA and electrification rates and transportation fuel types by state in the Western U.S. Interconnection. This data is needed by power grid operators and clean transportation planning authorities and industries to conduct their long-term planning studies. All the data is publicly available at \url{https://doi.org/10.5281/zenodo.13306893} \cite{acharya_2024_13306893}.

\end{enumerate}

\section{Methodology}
\label{sec:decarb_scenario}
We first present the modeling tools used to produce the decarbonization scenarios. Next, we detail how the IRA policies for the transportation sector are incorporated into the scenarios. Following that, we describe the integration of CCS technologies into the scenarios.

\subsection{Global Change Analysis Model}
\label{sec:gcam}
The Global Change Analysis Model is a widely employed, economically-driven, multi-sector dynamics model that simulates the interactions between human activities and natural systems. As a global model with detailed sectoral representations in the U.S., GCAM aims to simulate dynamic global transformations such as decarbonization and provide insights on their multi-sectoral impacts. Organizations such as the Intergovernmental Panel on Climate Change \cite{ipcc} and the U.S. Department of Energy have utilized GCAM extensively. GCAM and its ancillary models are calibrated from 1975 to the base year (currently set to 2015) and then run forward in time through 2100 with 5-year time-steps. We utilize GCAM-USA v6, which represents a sub-national economy and energy systems on 50 states plus the District of Columbia. We use a version that specifically represents the IRA policies and CCS technologies \cite{ou2023state}. In addition, GCAM-USA v6 integrates state-level socio-economic drivers, energy trends, and final energy services. We use the population and gross domestic product growth assumptions from Shared Socioeconomic Pathway-2 (SSP2) \cite{o2017roads}. The model adopts electricity technology costs from 2022 annual technology baseline data generated by National Renewable Energy Laboratory (NREL) \cite{EIA} and transportation cost assumptions, including EVs, primarily obtained from NREL’s electrification futures study \cite{jadun2017electrification}. 

\subsection{Modeling the IRA Policies in GCAM-USA}
\label{sec:IRA_transportation_gcam}
The Inflation Reduction Act (IRA) \cite{IRA} has provisions that impact clean transportation during the period 2023-2032. The IRA has provisions for incentivizing the buying of new and pre-owned clean fuel vehicles, building alternative fueling stations, domestic manufacturing of advanced/efficient clean fuel vehicles and their components, and the production of clean biofuels and sustainable aviation fuels.

A new clean vehicle credit (section 13401) allows credits up to \$7500 at the point of sale of clean vehicles from 2023-2032, including battery electric, plug-in hybrid, or fuel cell electric vehicles. EVs are eligible for half of the credit if their battery meets domestic assembly or manufacturing requirements. The other half of the credit is contingent upon a specific share of the minerals used in the battery being sourced from North American and other free trade countries. If the car meets the battery assembly and mineral sourcing requirements, a consumer can receive the full value of the tax credit provided that their income does not exceed the income eligibility threshold and that the sales price of the car does not exceed Manufacturer's Suggested Retail Price (MSRP) eligibility thresholds; $\$80,000$ for vans, SUVs, and pickup trucks and $\$55,000$ for general vehicles. 89\% of Americans meet the income requirement and GCAM-USA assumes that they would only purchase EVs that meet the MSRP threshold. Furthermore, we assume that the U.S. auto manufacturing sector will reorient itself so that half of EVs sold will meet the clean vehicle credit requirements by 2025 and all new EVs produced by 2030 will meet these requirements. 

Similarly, the pre-owned clean vehicle credit (section 13402) allows for credits up to $\$4,000$ for individuals meeting the income threshold and $\$40,000$ for businesses. Since section 13402 does not change the aggregated number of EVs in service (just their owners), GCAM-USA does not explicitly modeled this aspect. The clean commercial vehicle credit (section~13403) is modeled in GCAM-USA as a \$40,000 capital cost reduction for electric heavy duty freight trucks and a \$7,500 capital cost reduction for electric medium duty and light duty freight trucks.

The IRA’s alternative refueling property credit (section 13404) is an up to $\$1,000$ (for individuals) and $\$100,000$ (for businesses) property credit for building alternative refueling infrastructure, including electricity, ethanol, natural gas, hydrogen, and biodiesel in rural and low-income census tracts. Based on census data, 17.4\% of Americans live in counties that are either rural or low-income, so GCAM-USA models the \$1,000 property credit as a weighted average national subsidy of \$174 for capital infrastructure cost for EVs.

Sections 13201-13203 provide extended incentives for various biofuels, which was implemented as subsidies for biodiesel, cellulosic ethanol, Fischer-Tropsch (FT) biofuels, cellulosic ethanol with CCS technologies, and FT biofuels with CCS technologies in GCAM-USA. We assume that FT biofuel will primarily enter the jet fuel market first. As a result, FT biofuels will qualify for the aviation fuel credit.  

\subsection{Modeling CCS Technologies in GCAM-USA}
Carbon capture and storage technologies in GCAM-USA are modeled as options that can be applied to various processes, including those integrated with GHG-emitting generators like electricity production (e.g., coal-fired power plants with CCS), liquid fuel production, and fertilizer manufacturing, as well as standalone technologies like direct air capture CCS plants. The deatils about the CCS technologies in GCAM-USA is available in \cite{gcam_docuv7, gcam_docuv6}. The CCS-equipped technology competes with other technologies based on the relative cost, efficiency, and its ability to reduce carbon emissions if a carbon constraint (e.g., carbon tax) exists. Therefore, the adoption of CCS can vary across a range of future scenarios. GCAM-USA considers the cost of transporting and storing the captured CO$_2$ given the existing and potential future storage sites. Additionally, GCAM-USA captures regional differences in storage capacities, affecting the projected CCS deployment across states. Specifically for transportation, liquid fuel produced with CCS can be used by non-electric vehicles while ensuring a reduced carbon footprint.

\section{Experiment and Evaluation Metrics}
\label{sec:results}

We first present the numerical experiment of the scenarios used in this paper. Next, we discuss the parameters used for analyzing the impact of IRA policies and CCS technologies on transportation electrification. We study the impact of the IRA policies and CCS technologies using three parameters: i) transportation electrification rate, ii) transportation fuel mix, and iii) high-resolution spatio-temporal transportation charging loads. Furthermore, we describe how these parameters are helpful for transportation decarbonization authorities and power grid operators to achieve their decarbonization goal.

\subsection{Numerical Experiment}
\label{sec:pathways}
This study is based on three GCAM-USA scenarios with and without the IRA policies and CCS technologies. For all scenarios, the socioeconomic change assumptions are consistent with SSP2 \cite{o2017roads}. All scenarios include future climate impacts on heating and cooling demands are based on an RCP 8.5 radiative forcing pathway \cite{acharya2023weather}.We chose the RCP 8.5 climate pathway because it is most suitable for worst-case stress analysis of the grid. However, the rationale for selecting RCP 8.5 extends beyond the scope of this study and aligns with the approach used in other studies within the Grid Operations, Decarbonization, Environmental and Energy Equity Platform (GODEEEP) project \cite{godeeep}.

\subsubsection{Net-Zero (nz\_climate)}
This scenario simulates explicit targets of a clean electricity grid in the U.S. by 2035 and a fully net-zero GHG emissions economy by 2050. It does not include the IRA incentives. It also assumes that CCS technologies are unavailable. This decarbonization pathway was used in our earlier study \cite{acharya2023weather}.

\subsubsection{Net-Zero with CCS (nz\_ccs\_climate)}
Like the base Net-Zero scenario, this scenario includes a clean electricity grid in the U.S. by 2035 and a fully net-zero GHG emissions economy by 2050. It does not include the IRA incentives. It does however assume that CCS technologies are available.

\subsubsection{Net-Zero with IRA and CCS (nz\_ira\_ccs\_climate)}
This final scenario includes a clean electricity grid in the U.S. by 2035 and a fully net-zero GHG emissions economy by 2050. It does include the IRA incentives and assumes that CCS technologies are available. 

GCAM is calibrated to the base year 2015 and projects future scenarios with the aim of achieving a clean U.S. grid by 2035 and a net-zero U.S. economy by 2050. As a result, GCAM scenarios are designed to progress towards these goals, which means that discrepancies between the scenarios and actual observed rates of transportation electrification from 2020-current are expected.

\subsection{Transportation Electrification Rate}
\label{sec:electrification_rate_why}
The transportation electrification rate for a specific vehicle type measures the extent of electric energy usage within that type. It is defined as the ratio of the electric energy consumed by the vehicle type to the total energy consumed by that vehicle type within a given year. This rate can also be determined by examining the proportion of EVs within the total fleet of that vehicle type, regardless of the fuel types used. We analyze the transportation electrification rates of LDVs, MDVs, and HDVs separately. This segmented analysis is crucial for revealing nuanced relationships between vehicle types and electrification rates for at least two reasons. First, LDVs, MDVs, and HDVs differ significantly in their duty cycles, weight, and operational characteristics. LDVs, being lighter and designed for lower-duty applications, inherently demand smaller capacity batteries. This results in higher electrification rates, as LDVs address the common concern of range anxiety associated with EVs. In contrast, MDVs and HDVs, with their heavier weight and higher-duty requirements, present unique challenges in terms of battery capacity and charging requirements, which directly influences their electrification rates. Second, the separate analysis allows for a more granular examination of policies and their impact on different vehicle categories. States and regions often implement distinct regulations and incentives tailored to specific vehicle types based on factors such as emissions, transportation goals, and charging infrastructure development. Therefore, understanding the electrification rates for LDVs, MDVs, and HDVs independently provides insight into the diverse strategies employed by states to promote EV adoption within their jurisdictions. Furthermore, this segmented analysis is essential for stakeholders involved in EV infrastructure planning, deployment, and policy formulation. LDVs, MDVs, and HDVs serve distinct purposes in the transportation sector and their electrification rates are indicative of the feasibility and challenges associated with transitioning each vehicle category to its electric counterpart.

\subsection{Transportation Fuel Mix}
\label{sec:transport_fuel_mix_why}
The transportation fuel mix is essential for guiding strategic resource allocation and infrastructure development for fueling stations across a range of decarbonization scenarios. Different scenarios for the transportation sector results in different transportation fuel alternatives. Each transportation fuel has unique production method, cost, fuel supply challenges, raw material availability, and impact to the interconnected systems. For example, electricity consumed by EVs and fuel cell powered vehicles uniquely impact the demand on the power grid. Similarly, conventional Internal Combustion Engine (ICE) vehicles can be fueled in existing gas stations, while novel fuel powered vehicles require dedicated fuel distribution infrastructure (e.g., charging stations, hydrogen storage/fueling stations). To optimize fuel production and distribution facilities, an analysis of the transportation fuel mix for various scenarios and future years provides valuable insights for fuel suppliers, enabling them to strategically deploy resources based on regional consumption patterns and needs.

\subsection{Spatio-temporal Transportation Electric Loads}
\label{sec:spatio_temporal_why}
We downscale the annual transportation electricity demand into hourly charging loads across BAs in the Western U.S. Interconnection using the methodology described in our prior study \cite{acharya2023weather} and modified in Section~\ref{sec:downscaling_method_improvemnet}. Understanding the spatio-temporal distribution of charging loads across BAs helps power system operators identify peak and off-peak charging times and areas. This information allows them to plan resources and infrastructure to maintain supply-demand balance effectively. It also aids in production cost modeling and dynamic studies such as contingency analysis and transients. Furthermore, the spatio-temporal charging load profiles may inform power grid operators about the ratio of charging load to the total system load and help make informed decisions in various demand management schemes such as demand response and price-sensitive charging. On the charging side, the spatio-temporal charging loads informs charging station planning and deployment authorities to effectively allocate the charging infrastructure and increase their use time. 
Better understanding the range of spatio-temporal future charging load profiles can help increase the utilization factor of the charging stations and reduce the reversion to ICE vehicles.

\section{Impact of the IRA policies}
\label{sec:effect_of_IRA}

\subsection{Electrification Rate}
\label{sec:electrfication_rate_IRA}

\begin{figure}[!t]
    \centering
\includegraphics[width=1\columnwidth, clip=true, trim= 2mm 0mm 0mm 0mm]{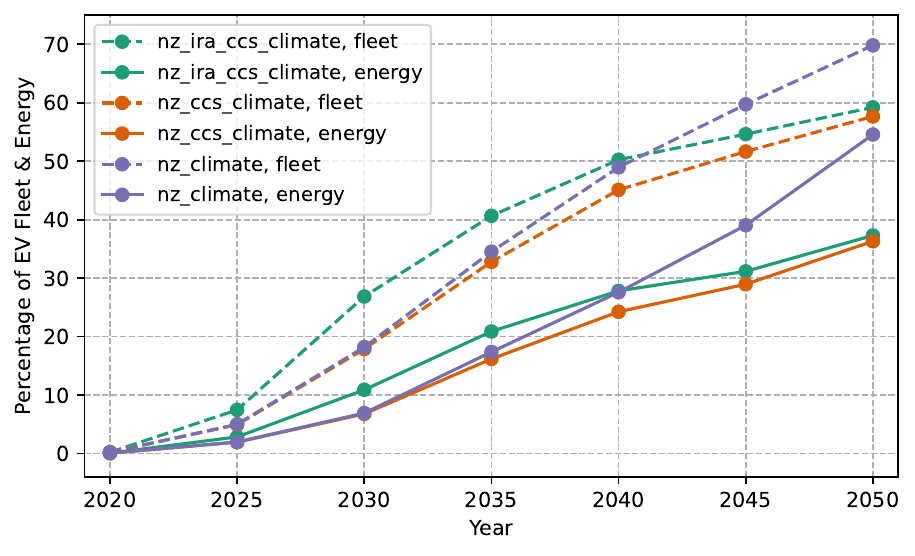}
    \caption{Combined transportation electrification rate for LDVs, MDVs, and HDVs: Percentage of EV energy and EV fleet across scenarios and over time in the Western U.S. Interconnection.}
    \label{fig:energy_fleet_ratio}
\end{figure}

\begin{figure*}[!t]
\centering
\subfigure[\label{fig:electric_fleet_nz_ira_ccs}]{
\includegraphics[width=0.58\columnwidth, clip=true, trim= 3mm 3mm 3mm 3mm]{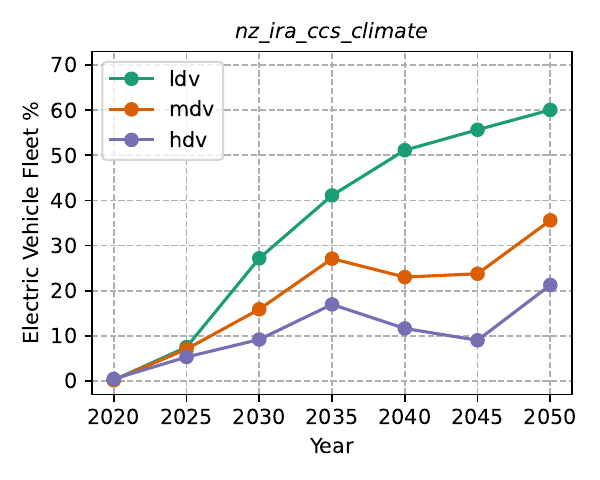}}
\hfill
\subfigure[\label{fig:electric_fleet_nz_ccs}]{
\includegraphics[width=0.58\columnwidth, clip=true, trim= 3mm 3mm 3mm 3mm]{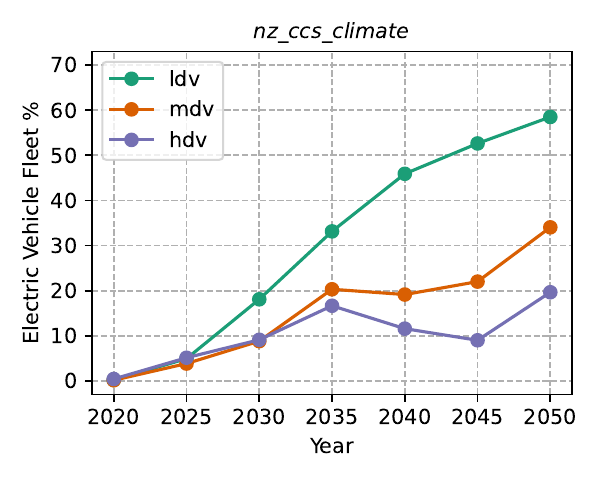}}
\hfill
\subfigure[\label{fig:electric_fleet_nz}]{
\includegraphics[width=0.58\columnwidth, clip=true, trim= 3mm 3mm 3mm 3mm]{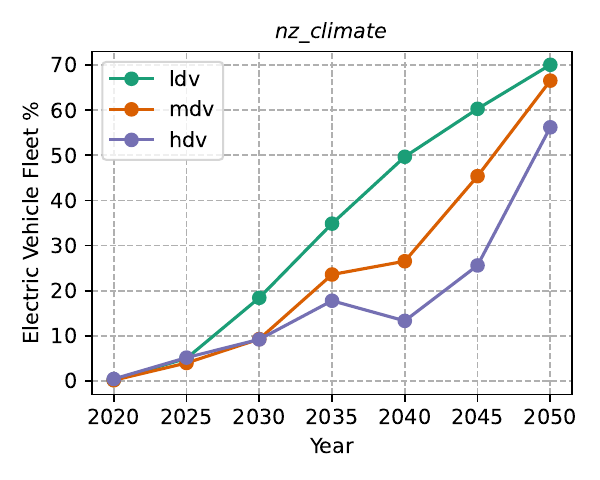}}
\caption{Transportation electrification, expressed as the percentage of electric vehicle fleet, for LDVs, MDVs, and HDVs over time and across scenarios in the Western U.S. Interconnection.}
\label{fig:electric_fleet}
\end{figure*}

Figure~\ref{fig:energy_fleet_ratio} illustrates the electrification rates of combined LDVs, MDVs, and HDVs within the Western U.S. Interconnection from 2020 to 2050 for each of our three scenarios described in Section~\ref{sec:pathways}. The electrification rates are quantified using two parameters. First, the proportion of EV energy, i.e., the ratio of electric energy consumed by LDVs, MDVs, and HDVs to the total energy consumed by LDVs, MDVs, and HDVs. Second, the proportion of EV fleet size, i.e., the ratio of number of electric LDVs, MDVs, and HDVs to the total number of LDVs, MDVs, and HDVs.

As depicted in Fig.~\ref{fig:energy_fleet_ratio}, the electrification rate of vehicles are identical and very small (0.08\% EV energy proportion and 0.23\% EV fleet size proportion) for all three scenarios in 2020 (which is a forward projected year in GCAM-USA), but start to diverge by 2025. This stems from the implementation of IRA policies starting in 2022. From 2025 onwards, scenarios incorporating the IRA policies exhibit higher electrification rates compared to those without. For instance, in 2025 the EV fleet size and EV energy proportions in \textit{nz\_ira\_ccs\_climate} scenario stand at 7.5\% and 4.9\% respectively, contrasting with 2.86\% and 1.96\% in \textit{nz\_ccs\_climate} scenario. This disparity persists into 2030 when the \textit{nz\_ira\_ccs\_climate} scenario has a 9\% elevation in EV fleet size fraction compared to the \textit{nz\_ccs\_climate} scenario. By 2035 the difference in EV fleet size fraction between \textit{nz\_ira\_ccs\_climate} and \textit{nz\_ccs\_climate} scenarios reduces to 8\%. This decrease in electrification rate between \textit{nz\_ira\_ccs\_climate} and \textit{nz\_ccs\_climate} scenarios in 2035 as compared to 2030 is because of the expiration of the IRA policies after 2032. The divergence in electrification rates between these scenarios diminishes after 2035 and they are nearly identical by 2050. However, the electrification rate in \textit{nz\_ira\_ccs\_climate} scenario is always higher than in \textit{nz\_ccs\_climate} scenario through 2050. This is for two reasons. First, the scenario with the IRA policies has a higher level of electrification to start with at 2035 as compared to the scenario without. Practically, the impact of IRA on electrification felt after 2032 can be due to the continuation of existing charging infrastructure and EVs manufactured during the IRA period. Second, the massive deployment of CCS after 2035 downplays the role of transportation electrification in achieving the net-zero economy in 2050, allowing the transportation electric rates to converge between the \textit{nz\_ira\_ccs\_climate} and \textit{nz\_ccs\_climate} scenarios. We explain the detailed role of CCS in transportation electrification in Section~\ref{sec:effect_of_ccs}.

Differences in the electrification rates across scenarios in Fig.~\ref{fig:energy_fleet_ratio} are evident in both EV fleet size and energy consumption. However, EV fleet fraction exhibits a higher value than EV energy fraction for any scenario and year. For instance, in 2030, \textit{nz\_ira\_ccs\_climate} scenario boasts a 9\% larger EV fleet fraction than \textit{nz\_ccs\_climate} scenario, while this difference reduces to 4\% in terms of EV energy fraction. Moreover, EV fleet fraction exhibits a steeper incline than EV energy fraction from earlier years (e.g., 2025) to later ones (e.g., 2030) for any given scenario. For instance, EV energy fraction in \textit{nz\_ira\_ccs\_climate} scenario increased from 10\% to 20\% between 2030 and 2035 while the EV fleet fraction increased from 20\% to 40\%. These trends mean the number of EVs are higher for the same amount of EV energy fraction as we go into the future, which is in-line with the increased fuel efficiency of EVs in the future.
 
Fig.~\ref{fig:electric_fleet} shows the individual LDV, MDV, and HDV EV fleet percentages for the three scenarios over time in the Western US Interconnection. To analyze the effect of IRA policies on individual EV types, we focus our analysis on Figs.~\ref{fig:electric_fleet_nz_ira_ccs} and \ref{fig:electric_fleet_nz_ccs}. We observe two key patterns. Firstly, the percentages of EVs in the LDV, MDV, and HDV fleets remain almost similar between \textit{nz\_ira\_ccs\_climate} and \textit{nz\_ccs\_climate} scenarios from the pre-IRA period (2020) to the early stages of the IRA (2025). However, the electrification rate in \textit{nz\_ira\_ccs\_climate} scenario increases faster than in \textit{nz\_ccs\_climate} scenario  in 2030 and 2035. This trend implies that the public initially takes time to respond to the benefits of the IRA policies regarding EV adoption but significantly increases their uptake afterward. Secondly, the IRA promotes significantly higher electrification rates for LDVs compared to MDVs and HDVs. For instance, in 2030, the LDV electrification rate in the \textit{nz\_ira\_ccs\_climate} scenario is approximately 10\% higher than in the \textit{nz\_ccs\_climate} scenario. In contrast, the MDV electrification rate is only about 7\% higher for the same year and scenarios. Moreover, the HDV electrification rate remains almost unchanged between these scenarios. These discrepancies suggest that the IRA policies have a limited impact on MDV and HDV electrification, possibly due to larger battery and charging infrastructure requirements and delays in manufacturing. In contrast, there are a growing number of LDV manufacturing companies. Furthermore, we model transitioning to mass transportation (e.g., rails and trains) from freight MDVs and HDVs in GCAM-USA to achieve net-zero US economy by 2050, which downplayed the need for HDV electrification in the IRA period as well.

\begin{figure*}[!t]
\centering
\subfigure[\label{fig:nz_ira_ccs_energy}]{
\includegraphics[width=0.58\columnwidth, clip=true, trim= 3mm 3mm 3mm 3mm]{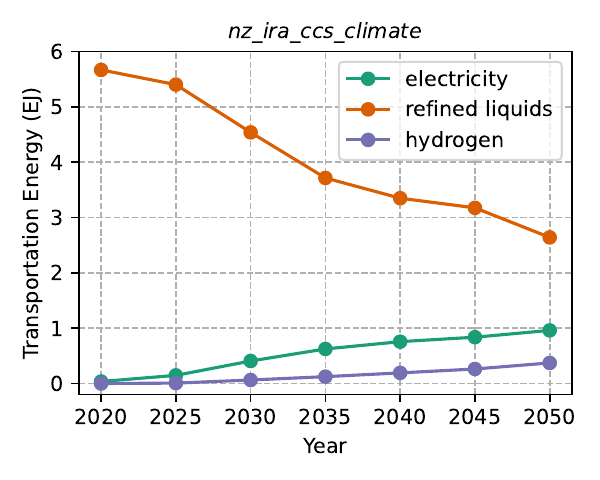}}
\hfill
\subfigure[\label{fig:nz_ccs_energy}]{
\includegraphics[width=0.58\columnwidth, clip=true, trim= 3mm 3mm 3mm 3mm]{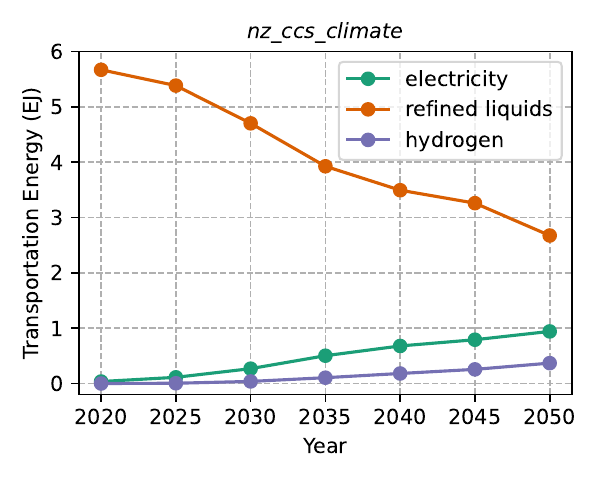}}
\hfill
\subfigure[\label{fig:nz_energy}]{
\includegraphics[width=0.58\columnwidth, clip=true, trim= 3mm 3mm 3mm 3mm]{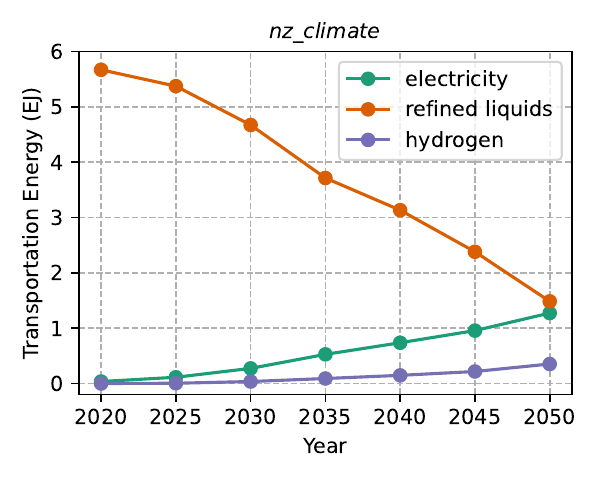}}
\caption{Transportation fuel mix over years for each of the three scenarios in the Western U.S. interconnection.}
\label{fig:fuel_mix}

\end{figure*}

\subsection{Transportation Fuel Mix}
\label{sec:transport_fuel_mix_IRA}
Fig.~\ref{fig:fuel_mix} shows the fuel mix in transportation sector in the Western U.S. Interconnection. We show transportation energy in exajoules (EJ) supplied by electricity, refined liquids, and hydrogen. Refined liquids include gasoline, diesel, and biofuels. To analyze the effect of the IRA policies in this section, we focus on \textit{nz\_ira\_ccs\_climate} in Fig.~\ref{fig:nz_ira_ccs_energy} and \textit{nz\_ccs\_climate} in Fig.~\ref{fig:nz_ccs_energy}. Similar to the transportation electrification rates in Figs.~\ref{fig:energy_fleet_ratio} and \ref{fig:electric_fleet}, the effect of IRA policies on transportation electrification and the use of hydrogen fuel in transportation sector due to IRA is minimal up to 2025 but both of them grow significantly in 2030 and 2035. For example, refined liquids energy in 2030 is 4.70 EJ in \textit{nz\_ccs\_climate} compared to only 4.53 EJ in \textit{nz\_ira\_ccs\_climate}, a 3.75\% decrease due to the implementation of the IRA policies. This decrease in GHG emitting fuels in the scenario with IRA policies is compensated by a 54\% increase in electrical energy (0.26 EJ in \textit{nz\_ccs\_climate} and 0.4 EJ in \textit{nz\_ira\_ccs\_climate}). Furthermore, hydrogen energy increases by 70\% under the IRA policies (0.037 EJ in \textit{nz\_ccs\_climate} to 0.063 EJ in \textit{nz\_ira\_ccs\_climate}). Although hydrogen energy increased by 70\% under IRA policies, it's consumption remains below 1\% till 2030 in the transportation sector. Similarly, the trend of decreasing refined liquid energy and increasing electric and hydrogen energy in the transportation sector continues under the IRA initiatives in 2035.
However, the fuel mix in \textit{nz\_ccs\_climate} and \textit{nz\_ira\_ccs\_climate} tends to become similar after 2035, leading to convergence by 2050. This is due to the cessation of IRA initiatives in 2032 and the adoption of CCS after 2035.

\begin{figure}[!t]
    \centering
\includegraphics[width=0.99\columnwidth, clip=true, trim= 0mm 1.1mm 0mm 0mm]{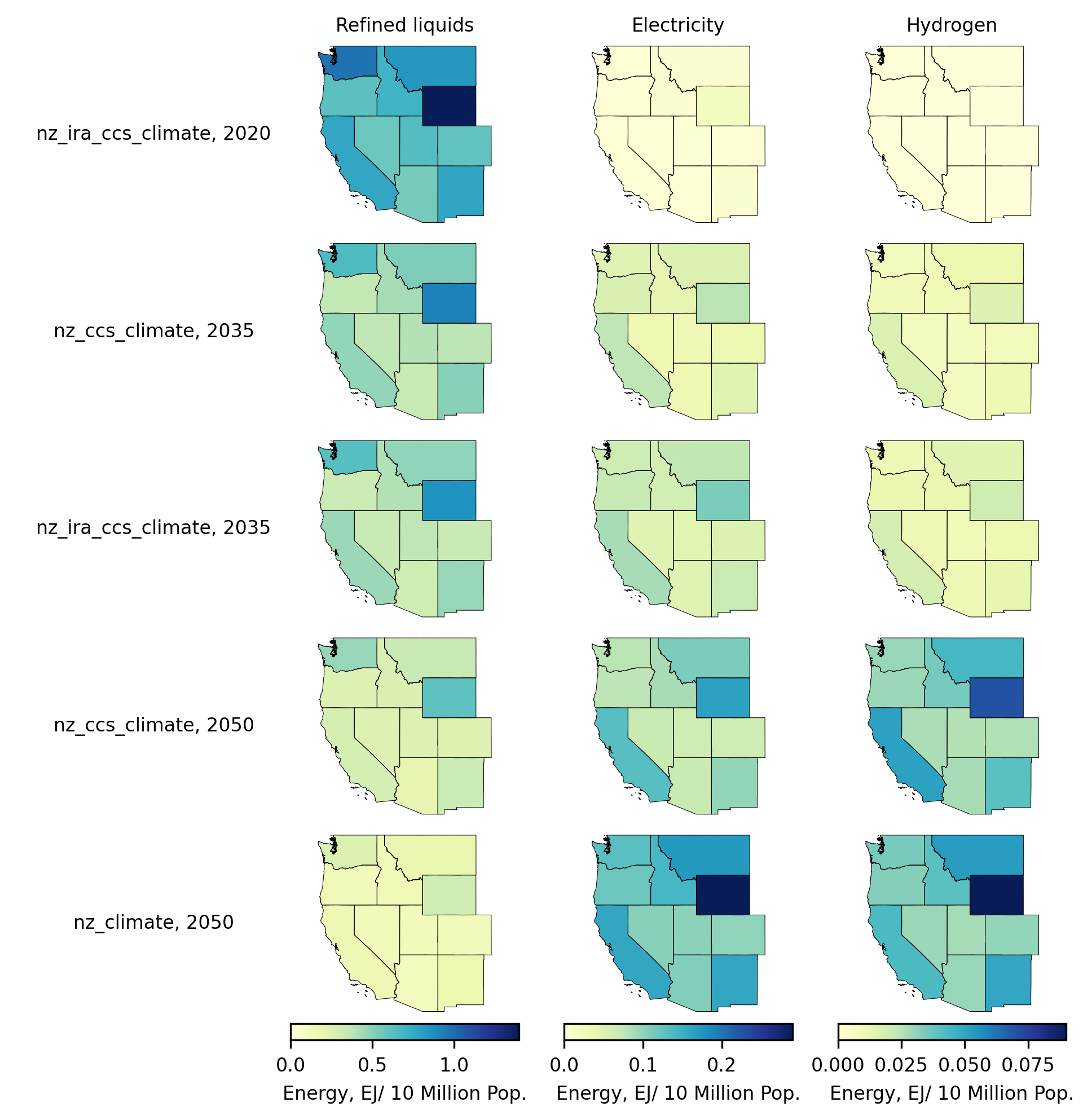}
    \caption{State-wise transportation fuel mix in the Western U.S. interconnection in 2035 and 2050 across the scenarios. The x-axis labels represent transportation fuel energy in EJ consumed by 10 million population. Different color scales are used for each fuel type to accurately capture and reflect the variation in their magnitudes.}
    \label{fig:fuel_mix_state_wise_heatmap}
\end{figure}

In Fig.~\ref{fig:fuel_mix_state_wise_heatmap}, we present a heatmap illustrating the fuel energy consumption in EJ by the transportation sector per 10 million people across states in the Western U.S. Interconnection. The \textit{nz\_ira\_ccs\_climate} scenario in 2020 serves as a benchmark for transportation fuel distribution. We examine the \textit{nz\_ccs\_climate} and \textit{nz\_ira\_ccs\_climate} scenarios in 2035 to assess the state-wise impact of the IRA policy on the transportation fuel mix. Two key observations emerge from Fig.~\ref{fig:fuel_mix_state_wise_heatmap} regarding the IRA's impact on transportation. First, refined liquids overwhelmingly dominate the transportation fuel mix in 2020 across all states, with electricity and hydrogen contributing minimally. For instance, in 2020, most states consume over 0.50 EJ per 10 million people in refined liquids, while electric energy and hydrogen consumption remains near zero. Second, states with less dense populations, such as Wyoming, exhibit a higher transportation energy consumption per capita compared to densely populated states like California and Washington. Due to the IRA's influence in 2035, the electric energy consumption in Wyoming increases from approximately 0.1 to 0.15 EJ per 10 million people from the \textit{nz\_ira\_ccs\_climate} scenario to the \textit{nz\_ccs\_climate} scenario. Conversely, the change in California during the same period and scenarios is minimal. These findings highlight the potential need for unique state-level policies to achieve a clean U.S. grid by 2035 and a net-zero U.S. economy by 2050.

\subsection{Spatio-temporal Load}
\label{sec:spatio-temporal_IRA}
\begin{figure}[!t]
\centering
\includegraphics[width=0.98\columnwidth, clip=true, trim= 0mm 1mm 0mm 0mm]{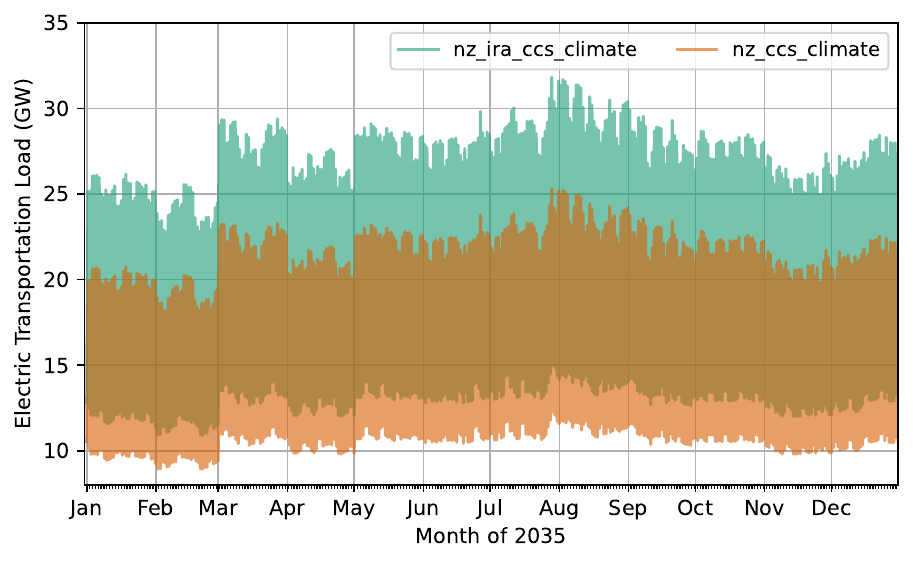}
\caption{Hourly time-series of total transportation electric load in the Western U.S. Interconnection for analyzing the impact of the IRA policies in 2035.}
\label{fig:time_series_ira}
\end{figure}

Fig.~\ref{fig:time_series_ira} presents an hourly time-series of the total electric transportation load, encompassing LDVs, MDVs, HDVs, aviation, rail, and maritime sectors, in the Western U.S. Interconnection for the year 2035. Please refer to Sections~ \ref{sec:downscaling_method} and \ref{sec:downscaling_method_improvemnet} in Appendix for the details on methodology for generating the time-series charging data. The scenarios \textit{nz\_ira\_ccs\_climate} and \textit{nz\_ccs\_climate} are examined to assess the impact of the IRA policies on the electric load profiles. Note that the profiles are generated using a diverse mix of charging strategies, including immediate charging (initiating as soon as the vehicle is parked), minimum power charging (maintaining a constant low power level throughout the dwelling period), and delayed charging (starting at the end of the dwelling period). These strategies are combined with various charging capacities, ranging from 3.7 kW to 500 kW. For detailed information on the exact mix used in generating the profiles, please refer to \cite{acharya2023weather}.

We make two key observations regarding the impact of the IRA policies on transportation loads based on Fig.\ref{fig:time_series_ira}. First, the impact of the IRA policies on transportation electric peak load is significantly higher than their impact on the proportion of electric energy and EV fleet sizes. For instance, the transportation electric peak load in 2035 is approximately 25\% higher in the \textit{nz\_ira\_ccs\_climate} scenario compared to the \textit{nz\_ccs\_climate} scenario. In contrast, the electric energy and EV fleet sizes are only about 8\% and 5\% higher, respectively, in the \textit{nz\_ira\_ccs\_climate} scenario as shown in Fig.\ref{fig:energy_fleet_ratio}. These differences indicate that the IRA's contributions vary significantly across various entities, including power grids, EV charging stations, and transportation fueling stations, in the context of transportation electrification. 
This is important because power grid operators are particularly concerned with the electric load profiles and peaks while charging station operators, planners, and clean vehicle fuel suppliers focus on the changes in transportation fuels or energy. Similarly, EV manufacturers are more interested in the impact on EV fleet size penetration due to the IRA policies.
Second, the difference between the maximum and minimum transportation loads increases markedly due to the IRA contribution. For example, this difference is 21 GW in the \textit{nz\_ira\_ccs\_climate} scenario compared to 16 GW in the \textit{nz\_ccs\_climate} scenario. This increased variability between valley and peak load poses challenges for power system operators, potentially requiring the over sizing of generators and the addition of storages \cite{wecc_load_variation}. Furthermore, such load variations over short time frames can lead to operational challenges, including equipment overloads and frequency and voltage fluctuations \cite{secchi2023smart}.

\begin{figure}[!t]
\centering
\includegraphics[width=0.99\columnwidth, clip=true, trim= 6mm 1mm 8mm 0mm]{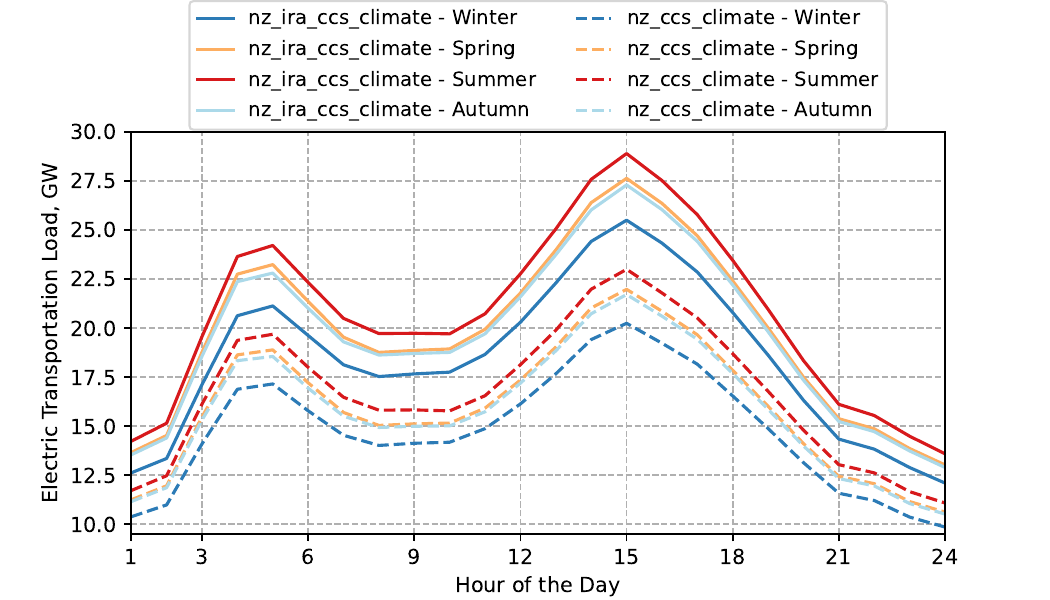}
\caption{Hourly average of electric transportation load (in UTC) across seasons in the Western U.S. Interconnection for analyzing the impact of IRA in 2035.}
\label{fig:season_ira}
\end{figure}

Fig.~\ref{fig:season_ira} illustrates the hourly average total transportation electric load profiles (in UTC time) in the Western U.S. Interconnection across seasons in 2035 for the \textit{nz\_ira\_ccs\_climate} and \textit{nz\_ccs\_climate} scenarios, highlighting the impact of the IRA policies on seasonal charging loads. Three primary observations are noted. Firstly, the peaks at approximately 5 AM and 3 PM UTC ($\approx$ 10 PM and 8 AM  local time in the Western U.S. Interconnection respectively) for all scenarios. 
Secondly, hourly loads are highest during summer and lowest during winter, with spring and autumn showing similar patterns across both scenarios. This seasonal variation in charging peak loads is primarily influenced by temperature fluctuations and vehicle mobility patterns. For instance, regions in the Pacific Northwest, such as Washington and Idaho, experience harsh winters, leading to reduced battery efficiency and increased charging demand. Conversely, southwestern states like California and Arizona, with their milder winter temperatures and extreme summer heat, see peak charging demands during the summer months \cite{yip2023highly}. Additionally,  given that regions with significant summer peaks also have higher EV penetration, the overall average charging demand is greater in summer than in winter. Notably, we do not 1) temporally downscale rails, aviation, and ships and 2) consider seasonality in temporal downscaling of MDVs and HDVs. Thus, the effect of seasonality in Fig.~\ref{fig:season_ira} is limited to LDVs. 
Thirdly, the difference between the maximum and minimum averaged charging peaks increases due to the IRA policies. For example, the difference between the summer peak and winter peak in 2035 is approximately 3.5 GW in the \textit{nz\_ira\_ccs\_climate} scenario, compared to about 2.7 GW in the \textit{nz\_ccs\_climate} scenario. The higher transportation peak load in summer may require that power system operators schedule additional generators and/or storage during the summer to maintain generator-load balance.
Overall, the contributions of the IRA policies and CCS technologies seem to remain similar across seasons.

\section{Impact of CCS Technologies}
\label{sec:effect_of_ccs}

\subsection{Electrification Rate}
\label{sec:electrification_rate_ccs}
We use Figs.~\ref{fig:energy_fleet_ratio} and \ref{fig:electric_fleet} to show the effect of CCS technologies on the electrification rate of LDVs, MDVs, and HDVs in the Western U.S. Interconnection.  In particular, we compare the \textit{nz\_ccs\_climate} and \textit{nz\_climate} scenarios and focus our analysis after 2035. As Fig.~\ref{fig:energy_fleet_ratio} shows, the EV energy and EV fleet percentages in the \textit{nz\_ccs\_climate} and \textit{nz\_climate} scenarios are almost same till 2030. After 2035 both EV energy and EV fleet percentages are higher in \textit{nz\_climate} scenario than in \textit{nz\_ccs\_climate} scenario. This trend continuously increases as we go to 2050.
For example, in 2050, the use of CCS technologies is associated with a 12\% smaller EV fleet (i.e., the electric fleet percentage is 58\% in \textit{nz\_ccs\_climate} scenario while it is 70\% in \textit{nz\_climate} scenario in Fig.~\ref{fig:energy_fleet_ratio}). In terms of percentage EV energy, the use of CCS technologies requires 53\% less EV energy in 2050 (i.e., the percentage EV energy is 36\%  in \textit{nz\_ccs\_climate} scenario compared to 55\% in the \textit{nz\_climate} scenario). Such a large reduction in EV energy and EV fleets in 2050 is due to the deployment of CCS. Notably, similar to the contribution of IRA on electrification rates in Section~\ref{sec:electrfication_rate_IRA}, the smaller percentage change in EV fleet compared to EV energy is due to the increase in the fuel efficiency of EVs in the future. 

Figure ~\ref{fig:electric_fleet} shows the percentages of LDV, MDV, and HDV EV fleets individually over years for the three scenarios. We use Figs.~\ref{fig:electric_fleet_nz_ccs} and \ref{fig:electric_fleet_nz} to gauge the the effect of CCS technologies on LDV, MDV, and HDV EV fleets. We make two  key observations. First, starting in 2035 the LDV, MDV, and HDV electrification rate is steeper without the use of CCS technologies. This enhanced electrification is needed to meet the net-zero economy goal in 2050. Second, the electrification of LDV, MDV, and HDV is significantly higher in the scenario without CCS available. For example, LDV, MDV, and HDV EV fleets in 2050 are 20\%, 32\%, and 38\% more in \textit{nz\_climate} scenario as compared to \textit{nz\_ccs\_climate} scenario, respectively.

\subsection{Transportation Fuel Mix}
\label{sec:fuel_mix_CCS}
We use Figs.~\ref{fig:nz_ccs_energy} and \ref{fig:nz_energy} to analyze the impact of CCS technologies on the transportation fuel mix needed to achieve the economy wide net-zero emission in 2050. We make two key observations about the effect of CCS technologies on transportation electrification in Figs.~\ref{fig:nz_ccs_energy} and \ref{fig:nz_energy}. 

Firstly, the effect of CCS technologies on transportation electrification is more evident after 2035 as the deployment of CCS technologies is in the later phase of decarbonization. This delay is due to the high initial costs and technical challenges of CCS deployments, and the need to first prioritize more immediately scalable solutions like renewable energy. Furthermore, CCS technologies allows the continuation of vehicles powered by GHG emitting fuels such as diesel and gasoline while still meeting the net-zero emission target in 2050. For example, in 2050, the transportation sector in the Western U.S. Interconnection consumes 1.5 EJ of refined liquid energy in \textit{nz\_climate} scenario compared to 2.7 EJ in the \textit{nz\_ccs\_climate} scenario. Note that the non-zero refined liquid energy in 2050 in the \textit{nz\_climate} scenario is due to the clean biomass liquids such as ethanol and biodiesel. Secondly, besides the higher level of refined liquid energy in \textit{nz\_ccs\_climate} scenario compared to \textit{nz\_climate} scenario, electric energy and hydrogen energy are also lower in the scenario with CCS technologies. For example, in 2050 electric energy is 35\% lower under CCS technologies (i.e., electric energy in \textit{nz\_ccs\_climate} scenario is 0.94 EJ compared to 1.27 EJ in \textit{nz\_climate} scenario). However, hydrogen energy in \textit{nz\_ccs\_climate} scenario is higher than \textit{nz\_climate} scenario. The difference in hydrogen energy in the two scenarios becomes smaller as we go to 2050. This trend of continuity in low-cost refined liquids, slow growth in electric energy, and increase in hydrogen energy are the compounded effects of CCS technologies on the transportation fuel mix. 

In Fig.~\ref{fig:fuel_mix_state_wise_heatmap} we utilize the \textit{nz\_ccs\_climate} and \textit{nz\_climate} scenarios for the year 2050 to examine the impact of CCS technologies across the states in the Western U.S. Interconnection. Two key observations regarding the state-level impact of CCS technologies on the transportation sector emerge from this analysis. First, mirroring the impact of the IRA policies on transportation discussed in Section~\ref{sec:transport_fuel_mix_IRA}, the changes in fuel mix due to CCS technologies are more pronounced in sparsely populated states than in densely populated ones. For instance, in 2050 electric energy consumption per 10 million population in Wyoming decreases from approximately 0.25 to 0.15 EJ when shifting from the \textit{nz\_climate} scenario to the \textit{nz\_ccs\_climate} scenario, whereas this change is minimal in California. Second, the use of CCS technologies enables a higher continuation of refined liquid usage in some states such as Washington and Wyoming under the \textit{nz\_ccs\_climate} scenario in 2050. This discrepancy may be attributed to the uneven deployment of CCS technologies across different states, as suggested by the distribution of existing CCS facilities in the U.S. \cite{ccs_us}.

\begin{figure}[!t]
\centering
\includegraphics[width=0.98\columnwidth, clip=true, trim= 0mm 1mm 0mm 0mm]{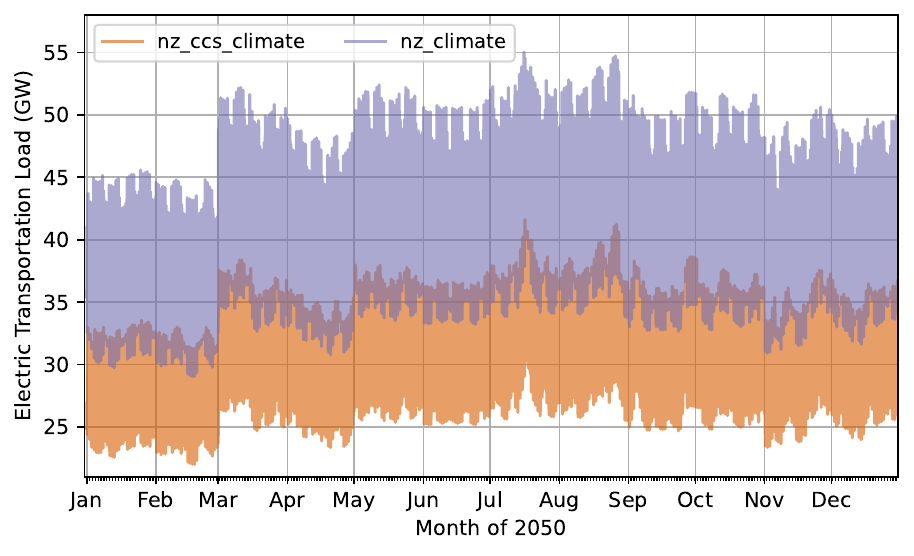}
\caption{Hourly time-series of total transportation electric load in the Western U.S. Interconnection for analyzing the impact of CCS technologies in 2050.}
\label{fig:time_series_ccs}
\end{figure}

\begin{figure}[!t]
\centering
\includegraphics[width=0.99\columnwidth, clip=true, trim= 6mm 1mm 8mm 0mm]{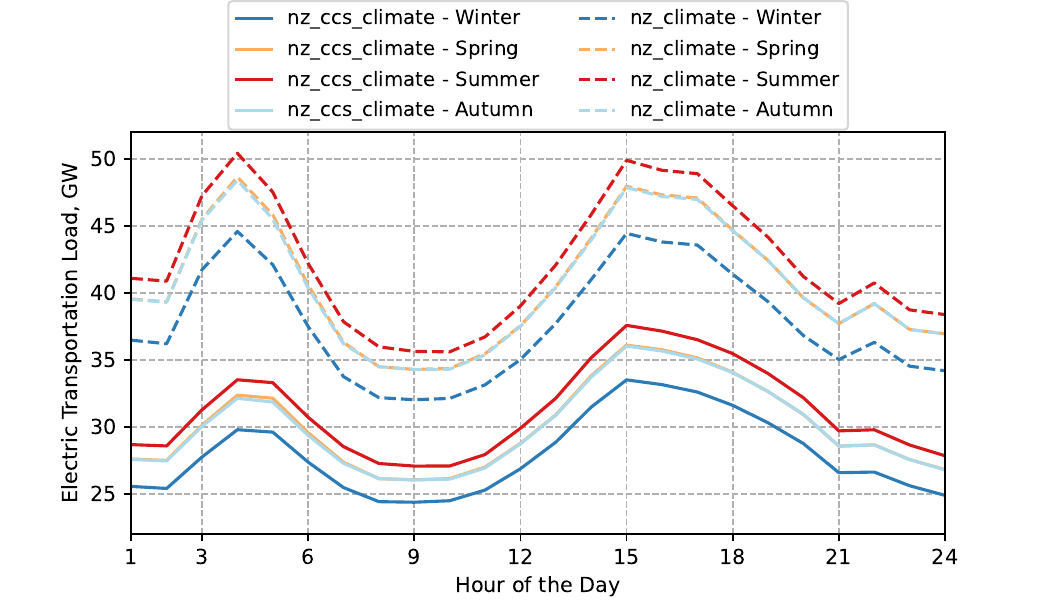}
\caption{Hourly average electric transportation load (in UTC) across seasons in the Western U.S. Interconnection for analyzing the impact of  CCS technologies in 2050.}
\label{fig:season_ccs}
\end{figure}

\subsection{Spatio-temporal Load}
\label{sec:spatio-temporal-ccs}

Figure~\ref{fig:time_series_ccs} shows an hourly time-series of the total transportation load, encompassing LDVs, MDVs, HDVs, aviation, rail, and ships, in the Western U.S. Interconnection in the year 2050. We use the \textit{nz\_ccs\_climate} and  \textit{nz\_climate} scenarios to examine the effect of CCS technologies on charging load profiles. Please refer to Section~\ref{sec:spatio-temporal_IRA} for the summary of the methodology of spatio-temporal load downscaling method. 

We make two key observations regarding the impact of CCS technologies on hourly charging load profiles based on Fig.~\ref{fig:time_series_ccs}. 
Firstly, the impact of CCS is different on peak charging load, EV fleet, and EV energy.
For example, the transportation electric peak in 2050 is $\approx 33\%$ higher in \textit{nz\_climate} scenario as compared to \textit{nz\_ccs\_climate} scenario, while it is $\approx 12\%$ and $\approx 18\%$ higher in terms of EV fleet and EV energy as shown in Fig.~\ref{fig:energy_fleet_ratio}, respectively. This trend suggests that the power grid operators observe smaller amount of peak load in 2050 with the deployment of CCS technologies, requiring potentially fewer power system upgrades. Similarly, EV manufactures and EV charging station operators observe lower EV demand and charging demand with CCS deployments, respectively. 
Secondly, the variability in the transportation charging load profiles decreases with the deployment of CCS technologies. For example, the difference between the maximum and minimum transportation charging load values in \textit{nz\_climate} scenario is  $26$ GW, while it is $20$ GW in \textit{nz\_ccs\_climate} scenario. Such variability in load profiles pose operational (e.g., frequency and voltage fluctuations) and resource allocation (e.g., new generators and transmission lines) challenges to power system operators\cite{wecc_load_variation,secchi2023smart}. However, with the CCS deployments, the load variations decrease, which potentially reduces the power grid operational stress and necessary infrastructure upgrades.

Figure ~\ref{fig:season_ccs} shows the hourly average total transportation load profiles (in UTC time) in the Western U.S. Interconnection across seasons in 2050 for the \textit{nz\_ccs\_climate} and \textit{nz\_climate} scenarios. This figure highligts the impact of CCS technologies on seasonal charging loads. We make two key observations in Fig.~\ref{fig:season_ccs}. Firstly, the transportation charging peaks at approximately 4 AM and 3 PM UTC time ($\approx$ 9 PM and 8 AM Western U.S. local time respectively) are highest in summer and the lowest in winter, while peaks in other seasons falls in between and are alike. Similar to Section~\ref{sec:spatio-temporal_IRA}, the seasonal variation in charging load can be influenced by temperature fluctuations, vehicle mobility patterns, and dominance of EV penetration in southern states in the Western U.S. Interconnection.
Secondly, the variation of charging peaks across seasons is smaller due to CCS deployments. For example, the difference between summer and winter peaks in \textit{nz\_ccs\_climate} scenario is $4$ GW while it is $\approx 6$ GW in \textit{nz\_climate} scenario. Understanding seasonal peak load variation is essential for power system operators to ensure grid stability and reliability by optimizing resource allocation, planning demand response, and managing costs effectively during high-demand periods \cite{wecc_load_variation}. It also aids in the seamless integration of renewable energy and long-term infrastructure planning. This smaller variability in transportation charging peak loads with the CCS technologies implies lesser need of additional resource scheduling (e.g., generator, battery storage) for the power system operators to meet the transportation peak loads in summer. 

The average seasonal load profile in 2050 in Fig.~\ref{fig:season_ccs} is different than that in Fig.~\ref{fig:season_ira} in 2035 because of dynamic LDV charging behaviors in future. For instance, we use 30\% immediate (\emph{min\_delay}) and 70\% constant minimum power (\emph{load\_level}) charging, while they are 80\% and 20\%, respectively in 2035. This charging scenario is to reflect significant home-based LDV charging in later years. Please refer to our earlier study \cite{acharya2023weather} for detailed information on charging strategies and their mix.

\section{Discussion and Conclusions}
\label{sec:conclusion}
In this paper, we found that the interactions between IRA policies and CCS technologies on  transportation electrification metrics, especially prior to 2035 were minimal. The impact of the IRA policies on the transportation sector tend to concentrate from 2025 to 2035 and the impact of CCS technologies  from 2035 to 2050.    
The electrification rates in \textit{nz\_ira\_ccs\_climate} and \textit{nz\_ccs\_climate} scenarios converge after 2035, leading to similar outcomes by 2050, as illustrated in Fig.~\ref{fig:energy_fleet_ratio}. This convergence is attributed to the dominance of CCS technologies post-2035 and the expiration of the IRA policies in 2032.

This paper concludes that the U.S. IRA and CCS technologies significantly impact transportation electrification patterns in the Western U.S. Interconnection. Using the U.S. version of the Global Change Analysis Model (GCAM-USA), we modeled decarbonization scenarios targeting a clean U.S. grid by 2035 and a net zero US economy by 2050. Our analysis reveals that the IRA policies boost transportation electric energy by 54\% and hydrogen fuel adoption by 70\% by 2035, with a primary impact on light-duty vehicle electrification. After 2035, the adoption of CCS technologies reduces the transportation sector's electric energy consumption by 35\% by 2050. While it appears that it might not be critical to consider CCS technologies in near term resource adequacy and reliability plans (2035), infrastructure investments for post 2035 are being decided now. Despite the substantial anticipated contributions of CCS to achieving net zero goals by 2050, their adoption and deployment in the U.S. remain slow. Consequently, there remains a large uncertainty in the electrification of transportation. This uncertainty impacts investment decisions for both EV charging infrastructure authorities and EV manufacturers.

\section*{Acknowledgment}
This research was supported by the  Grid Operations, Decarbonization, Environmental and Energy Equity Platform (GODEEEP) Investment at Pacific Northwest National Laboratory (PNNL). PNNL is a multi-program national laboratory operated for the U.S. Department of Energy (DOE) by Battelle Memorial Institute under Contract No. DE-AC05-76RL01830. 

\appendix
\subsection{Spatio-temporal Charging Load Downscaling Methodology}
\label{sec:downscaling_method}
Here, we provide a summary of our methodology presented in \cite{acharya2023weather} for downscaling the annual electric energy in GCAM to hourly time-series charging loads across BAs in WECC. This methodology utilizes state-level yearly electric energy data for LDVs, MDVs, and HDVs from the GCAM-USA model to develop hourly electric charging loads across BAs. However, due to the lack of temporal downscaling data, the methodology assumes a flat electric charging load for rail, ships, and aviation. Thus, only spatial downscaling is performed for rail, ships, and aviation electric loads. Additionally, the spatio-temporal downscaling method models the charging loads separately for each month, resulting in sudden jumps in load profiles during month transitions. Improvements in the load downscaling methodology in Section~\ref{sec:downscaling_method_improvemnet} include incorporating en-route charging capabilities for MDVs and HDVs.

\subsection{Improvements to Charging Load Downscaling Method}
\label{sec:downscaling_method_improvemnet}

\begin{figure}[!t]
    \centering
\includegraphics[width=1\columnwidth, clip=true, trim= 42mm 20mm 30mm 50mm]{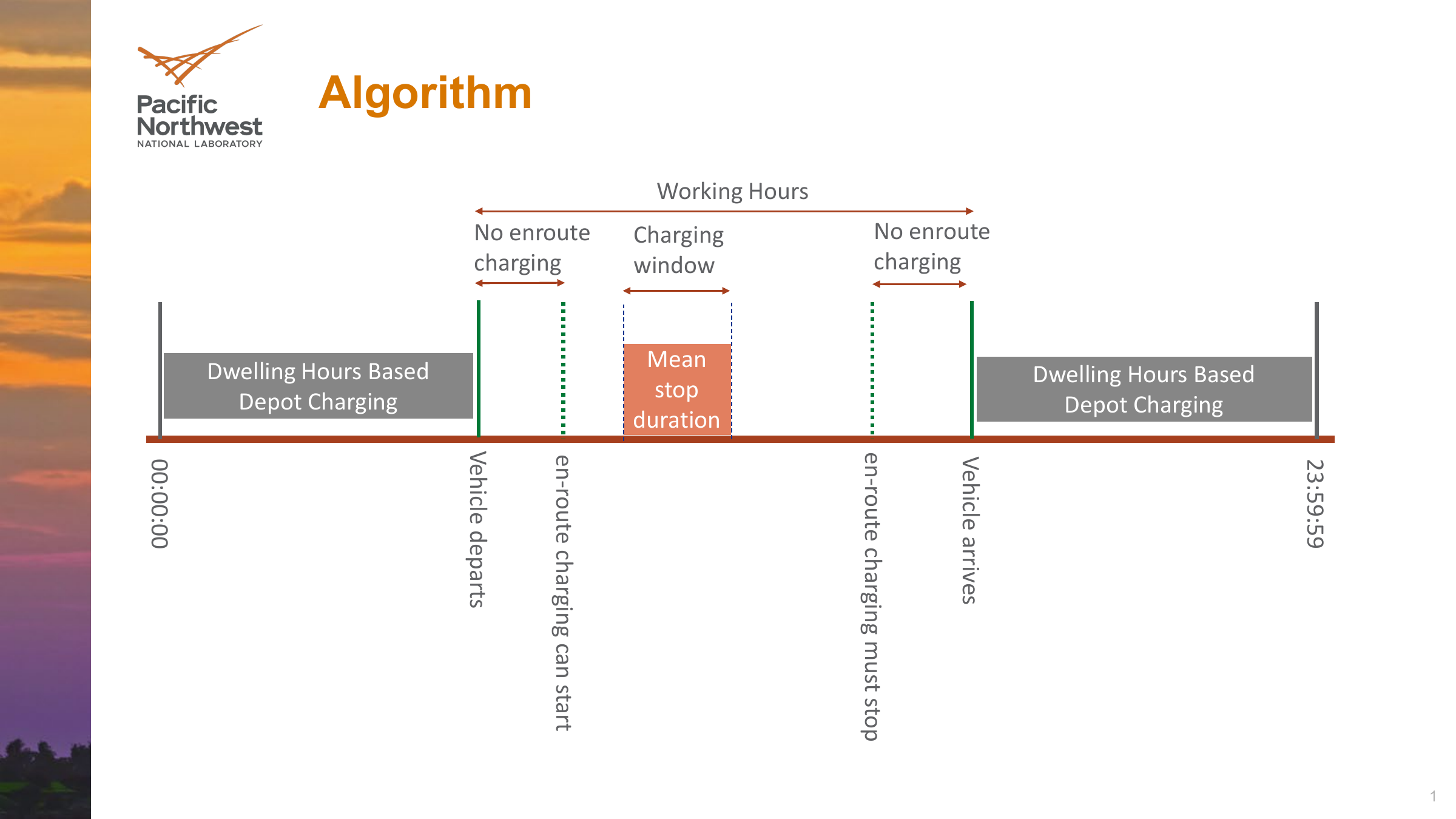}
    \caption{Schematic diagram for en-route MHDV charging on top of depot-based charging presented in \cite{acharya2023weather}.}
    \label{fig:en-route_method}
\end{figure}

Figure~\ref{fig:en-route_method} shows the schematic of en-route and opportunity charging for Medium-and-Heavy Duty Vehicles (MHDVs). Although depot charging constitutes a majority (e.g., 87\% \cite{muratori2022perspectives}) of the MHDV charging, en-route and opportunity charging will gain momentum as MHDV electrification deepens. The en-route charging in Fig.~\ref{fig:en-route_method} allows for charging during stops including loading/unloading and breaks. We assume that MHDVs do not charge right after leaving the depot or immediately before arriving at the depot. Furthermore, to avoid the bias in sampling charging window, we randomize the charging window between \textit{\textquotedblleft en-route charging can start"} and \textit{\textquotedblleft en-route charging must stop"}. Our study found that the incorporation of the en-route charging reduces the MHDV charging peaks as compared to only depot-based charging. 

\subsection{Data Availability}
\label{sec:data_code}
We provide three datasets in \url{https://doi.org/10.5281/zenodo.13306893}\cite{acharya_2024_13306893} relevant to our study:
\subsubsection{Spatio-temporal Charging Load Data} This dataset includes charging load data by Balancing Authorities within the Western U.S. Interconnection, corresponding to the \textit{nz\_ira\_ccs\_climate}, \textit{nz\_ccs\_climate}, and \textit{nz\_climate} scenarios. The data are available for the year 2035 for \textit{nz\_ira\_ccs\_climate} scenario, for 2035 and 2050 for \textit{nz\_ccs\_climate} scenario, and for 2050 for \textit{nz\_climate} scenario.
 \subsubsection{State-Level Electrification Rates} This dataset provides electrification rates at the state level, expressed as the percentage of EV energy use and fleet size for LDVs, MDVs, and HDVs under the \textit{nz\_ira\_ccs\_climate}, \textit{nz\_ccs\_climate}, and \textit{nz\_climate} scenarios. Data are available for the years 2020 through 2050, with a 5-year time step.
\subsubsection{State-Level Transportation Fuel Mix} This dataset details the state-level distribution of transportation fuel types, including hydrogen, electricity, and refined liquids, across the \textit{nz\_ira\_ccs\_climate}, \textit{nz\_ccs\_climate}, and \textit{nz\_climate} scenarios, spanning the years 2020 to 2050 in 5-year increments.

\bibliographystyle{IEEEtran}
\bibliography{ref}
\end{document}